\documentclass[sn-mathphys,Numbered]{sn-jnl}

\usepackage{graphicx}%
\usepackage{multirow}%
\usepackage{amsmath,amssymb,amsfonts}%
\usepackage{amsthm}%
\usepackage{mathrsfs}%
\usepackage[title]{appendix}%
\usepackage{xcolor}%
\usepackage{textcomp}%
\usepackage{manyfoot}%
\usepackage{booktabs}%
\usepackage{algorithm}%
\usepackage{algorithmicx}%
\usepackage{algpseudocode}%
\usepackage{listings}%

\theoremstyle{thmstyleone}%
\theoremstyle{thmstyletwo}%

\theoremstyle{thmstylethree}%

\begin{document}

\title[Article Title]{DyFeO\textsubscript{3} electrode material with ultra-wide voltage window for aqueous symmetric supercapacitors}


\author {\fnm{Mohasin} \sur{Tarek}}

\author {\fnm{Ferdous} \sur{Yasmeen}}

\author* {\fnm{and M. A.} \sur{Basith*}}
\email{mabasith@phy.buet.ac.bd}

\affil
{Nanotechnology Research Laboratory, Department of Physics, Bangladesh University of Engineering and Technology, Dhaka-1000, Bangladesh}




\abstract{Aqueous supercapacitors (SCs) encounter limitations in operational voltage and energy density due to the low decomposition voltage of water. Here, we fabricate aqueous symmetric supercapacitors (ASSCs) employing DyFeO\textsubscript{3} as an electrode material. This hybrid SC in a 0.5 M Na\textsubscript{2}SO\textsubscript{4} aqueous electrolyte exhibits a significantly high working voltage of 2.5 V, with an energy density of 41.81 Wh kg\textsuperscript{-1} at a power density of 1250 W kg\textsuperscript{-1}, maintaining 94\% capacitance retention after 5000 cycles. By incorporating 20\% volume of acetonitrile with water in the electrolyte, we extend the potential window to 3.1 V, with an energy density of 84.43 Wh kg\textsuperscript{-1} at a power density of 1550 W kg\textsuperscript{-1}. The as-fabricated ASSC shows promising stability during a 300-hour float voltage test with almost intact capacitance retention and Coulombic efficiency. For the first time, our study unveils the potential of porous DyFeO\textsubscript{3} as an electrode material for advancing ASSCs, featuring an unprecedented ultra-wide voltage window, along with significantly large energy and power densities. 

}




\maketitle

\section{Introduction}\label{sec1}
Electrochemical supercapacitors (SCs) are emerging as promising alternatives for sustainable and reliable energy storage devices \cite{doi:10.1126/science.aan8285}. Their advantages, including high power density, long cycle life, fast charge-discharge rates, wide operating temperature range, and safety, make them particularly compelling when compared to traditional batteries \cite{frith2023non,albertus2018status,10.1063/1.3360354}. Nevertheless, the limited energy densities, typically around 1-10 Wh kg\textsuperscript{-1}, observed in commercially available supercapacitors significantly constrain their broad application \cite{simon2008materials,wang2012review}. Consequently, there is a pressing need for extensive research involving various parameters in particular electrode materials and electrolytes to enhance energy densities and associated factors like cost, safety, and lifetime to unlock the full potential of SCs \cite{doi:10.1126/science.aan8285}. In the present context, maximizing energy density (E) is achieved through the effective expansion of the voltage window (V), leveraging the square proportionality between V and E for a specific capacitance. Notably, the stability window of aqueous supercapacitors (SCs) is restricted by the narrow stability range of water, approximately 1.23 V \cite{bu2019low,huang2023ultrahigh}. In pursuit of a significant expansion of stability windows for constructing SCs, the careful selection of electrode materials and the judicious choice of electrolytes emerge as key factors \cite{C4EE03229B,wang2012review,pal2019electrolyte}.

To enhance the electrochemical energy storage capabilities of electrode materials, various strategies, including doping, nanocomposite formation, molecular cross-linking, intricate structural design, and chemical modifications, are employed \cite{xu2022aqueous,bu2019low,wang2019wide,wang2020hierarchical,wang2012review}. While employing these strategies enhances electrochemical energy storage capabilities, it is crucial to acknowledge that they can often introduce complexity, reduce reliability, and increase costs. Particularly, in designing asymmetric SC, a common strategy to enhance voltage windows involves employing diverse complex electrode materials for the cathode and anode \cite{huang2022wide}. This approach, coupled with intentional imbalances in mass and charge between the electrodes, amplifies the intricacy and cost associated with both the fabrication and functionality of the SCs \cite{huang2022wide}. Therefore, the adoption of a unified, perovskite-like electrode material for both the cathode and anode, capable of demonstrating a hybrid charge storage mechanism involving simultaneous utilization of electrochemical double-layer capacitance and pseudocapacitance, may prove advantageous \cite{cao2021recent, mefford2014anion}. 


Notably, perovskite oxides, with the formula ABO\textsubscript{3} (where A represents a lanthanide or alkali earth element, and B denotes a transition metal), have been a focal point in electrode research due to their structural stability, compositional adaptability, and inherent oxygen vacancies \cite{cao2021recent, mefford2014anion}. Among various perovskite oxide materials, dysprosium orthoferrite (DyFeO\textsubscript{3}), crystallizing in a perovskite orthorhombic structure \cite{afanasiev2021ultrafast}, has been extensively studied for its potential applications in magnetic data storage, spintronics devices, and as a multiferroic material \cite{cao2016tuning, biswas2022role,zeng2020rare}. Despite these well-explored areas, DyFeO\textsubscript{3}'s unique electronic structure—characterized by a suitable band gap, diverse oxidation states of the constituent elements, and inherent oxygen vacancies—designates it as a promising candidate for applications beyond spintronics, particularly in electrochemical energy storage systems \cite{cao2021recent}. The diverse oxidation states of the constituent elements, Dy and Fe, may enable redox reactions, particularly in applications involving charge transfer within electrochemical systems \cite{cao2021recent}. Notably, the inherent oxygen vacancies can promote more efficient redox reactions at the electrode-electrolyte interface, enhance the material's surface reactivity, and serve as defect sites that facilitate ion diffusion within the material, potentially influencing the specific capacitance of the DyFeO\textsubscript{3} electrode material \cite{cao2021recent,hussain2023overview,doi:10.1021/cs500606g,shafi2018enhanced}. Moreover, the porous structure of DyFeO\textsubscript{3}, achieved through a controlled fabrication route, offers numerous electroactive sites, thereby enhancing the charge storage capacity in SCs \cite{cao2021recent}. This characteristic not only improves the material’s suitability for SC applications but also helps mitigate water-splitting reactions at elevated voltages, consequently expanding the electrochemical stability window (ESW) of the electrolyte solution. Hence, the distinctive characteristics of DyFeO\textsubscript{3} have strongly motivated us to employ it for the first time as electrode material in high-performance SC applications, demonstrating its potential for advancing electrochemical energy storage devices.

Apart from the electrode material, the choice of a suitable electrolyte is a crucial factor that significantly impacts the stability and overall effectiveness of SCs \cite{pal2019electrolyte,zhong2015review}. Currently, nonaqueous electrolytes are extensively used in prominent energy storage systems for portable electronics and automotive purposes due to their wide operating voltage range and high energy density; however, challenges such as flammability, significant economic costs, and environmental concerns restrict their widespread applicability \cite{zhong2015review,park2019rational}. In contrast, aqueous electrolytes present a promising alternative, offering cost-effectiveness, high ionic conductivity, intrinsic safety, and environmental compatibility \cite{xu2022aqueous,pal2019electrolyte,bu2019low,huang2023ultrahigh}. 
Considering the advantageous features of aqueous electrolyte solutions, a 0.5 M Na\textsubscript{2}SO\textsubscript{4} aqueous solution is chosen as the electrolyte in this investigation. This choice is based on its neutral pH, non-corrosive nature, small hydrated ion size, excellent conductivity, and cost-efficiency \cite{zhong2015review,fic2012novel,chang2014green}. 

To further expand the ESW of aqueous electrolytes, various approaches have been explored, such as adjusting the pH, increasing electrolyte concentration, modifying the potential of zero voltage of electrodes, introducing redox-active additives, balancing electrode masses, and passivating electrode surfaces \cite{bu2019low,xu2022aqueous,chun2015design}. However, these strategies, whether employed individually or collectively, have demonstrated only modest success in expanding the ESW. For instance,
employing a super-concentrated electrolyte with 21 M lithium bis(trifluoromethanesulfonyl)imide (LiTFSI) resulted in a “water-in-salt” (WIS) system, extending the electrochemical stability window (ESW) to 2.4 V \cite{hasegawa2016hierarchically}. Another study utilized a 17 M sodium perchlorate aqueous electrolyte, achieving an ESW of 2.3 V with a Coulombic efficiency of 66\% after 10,000 cycles \cite{bu2019low}. Conversely, incorporating organic solvents such as acetonitrile, ethanol, and ethylene glycol as additives in a WIS electrolyte has shown promise in enhancing the ESW of electrolyte solutions \cite{huang2023ultrahigh,dou2018safe}. Interestingly, the incorporation of organic solvents to increase the ESW, primarily by lowering salt concentration and decreasing the activity of free water molecules, proves to be a simple and cost-effective technique \cite{huang2023ultrahigh}. 



Therefore, in addition to a pure aqueous 0.5 M Na\textsubscript{2}SO\textsubscript{4} electrolyte, we developed an aqueous-dominant and dilute salt-containing solid electrolyte interface (SEI) system. This involved incorporating only 10\% and 20\% volumetric percent acetonitrile (AN) with water as the solvent medium for the 0.5 M Na\textsubscript{2}SO\textsubscript{4} electrolyte. Subsequently, we fabricated coin cell supercapacitors utilizing DyFeO\textsubscript{3} nanoparticles as the electrode material for both the cathode and anode, with electrolyte solutions of 0.5 M Na\textsubscript{2}SO\textsubscript{4}(aq.) and 0.5 M Na\textsubscript{2}SO\textsubscript{4}(aq.)/20\%AN. Notably, the diverse oxidation states, porous structure, and inherent oxygen vacancies of the DyFeO\textsubscript{3} electrode material, in combination with the 0.5 M Na\textsubscript{2}SO\textsubscript{4} aqueous electrolyte solution, collectively mitigate water splitting. This breakthrough extends the ESW to 2.5 V, marking to the best of our knowledge, the first instance of such an achievement for a hybrid aqueous symmetric supercapacitor. Furthermore, the electrolyte solution with a 20\% volumetric organic additive, AN, i.e.,0.5 M Na\textsubscript{2}SO\textsubscript{4}(aq.)/20\%AN, remarkably expanded the ESW to 3.1 V. Thus, the as-fabricated aqueous electrolyte-dominant hybrid symmetric SC demonstrates an impressive energy density of 84.43 Wh kg\textsuperscript{-1} at a power density of 1550 W kg\textsuperscript{-1}, coupled with excellent rate capability and cycle performance. These results surpass the integrated performance of any previously reported aqueous electrolyte-dominated symmetric SC. 



\section{Results}\label{sec2}
\begin{figure}[h]%
\centering
\includegraphics[width=1\textwidth]{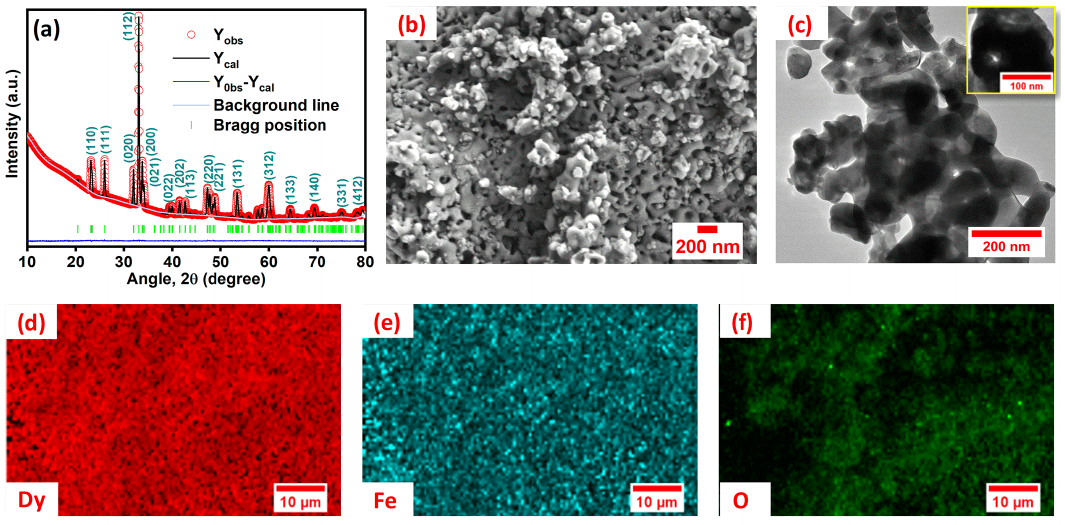}
\caption{\textbf{Structural and morphological characterization of the DyFeO\textsubscript{3} nanoparticles}. \textbf{a,} The Rietveld-refined powder X-ray diffraction (XRD) pattern indicates the crystalline nature of DyFeO\textsubscript{3} in the orthorhombic phase with the \textit{pnma} space group. \textbf{b,} FESEM image reveals the porous structure of DyFeO\textsubscript{3} nanoparticles. \textbf{c,} TEM image also confirms the porous structure of DyFeO\textsubscript{3}, as observed in FESEM imaging; the inset of (f) provides a closer view of the pore within a nanoparticle. \textbf{d-f,} Elemental mapping images exhibit the distribution of Dy, Fe, and O atoms in DyFeO\textsubscript{3} nanoparticles. }\label{Fig_1}
\end{figure}

The as-synthesized DyFeO\textsubscript{3} crystallizes in an orthorhombic structure with the space group \textit{pnma}\cite{10.1063/1.3360354,doi:10.1021/jp109313w}, as confirmed by the Rietveld refinement of the powder XRD pattern presented in Fig. 1. The crystallographic parameters and reliability factors are provided in Supplementary Table S3. The calculated crystallinity (93\%) confirms the formation of highly crystalline DyFeO\textsubscript{3} particles. The successful formation of DyFeO\textsubscript{3} nanoparticles with a porous structure is evident in the FESEM and TEM images, as depicted in Fig. 1(b) and 1(f), respectively. Most of the pore sizes estimated from the electron microscopy images range from approximately 10 to 45 nm (ESI Fig. S5(a)). Subsequently, EDX analysis, as illustrated in supplementary Fig. S5(b), was conducted to confirm the presence of each element in the synthesized material. The experimentally determined mass and atomic percentages of the constituent elements were found to closely match their theoretical values, as detailed in supplementary Table S1. Following the EDX analysis, the elemental mapping of DyFeO\textsubscript{3} in Fig. 1(c-e) reveals a uniform distribution of Dy, Fe, and O throughout the material, confirming its homogeneity.



\begin{figure}[h]%
\centering
\includegraphics[width=1\textwidth]{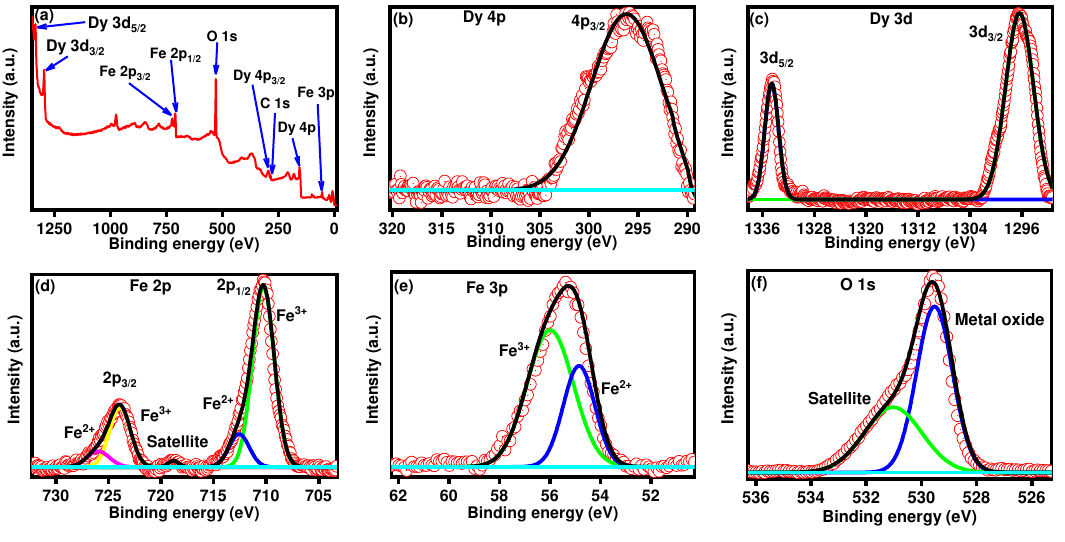}
\caption{\textbf{The chemical state analysis of DyFeO\textsubscript{3} nanoparticles through XPS.} \textbf{a,} The full survey spectrum reveals distinct peaks corresponding to the oxidation states of Dy\textsuperscript{4+}, Dy\textsuperscript{3+}, Fe\textsuperscript{2+}, Fe\textsuperscript{3+}, and O\textsuperscript{2-}. The identified oxidation states may enhance the electrochemical charge storage capability of DyFeO\textsubscript{3} as electrode material. High-resolution XPS spectra of \textbf{b,} Dy 4p, \textbf{c,} Dy 3d, \textbf{d,} Fe 2p, \textbf{e,} Fe 3p, and \textbf{f} O 1s further detail the specific binding energies associated with each element. The high-resolution XPS spectrum of the O 1s region \textbf{f,} reveals a satellite peak at 531.05 eV, indicating the presence of oxygen vacancies in DyFeO\textsubscript{3} nanoparticles. These vacancies have the potential to enhance the electrochemical performance of DyFeO\textsubscript{3} electrode material, as discussed in the text.}\label{Fig_2}
\end{figure}

X-ray photoelectron spectroscopy (XPS) was employed to analyze the chemical state of synthesized DyFeO\textsubscript{3} nanoparticles, as illustrated in Fig.2. The survey spectrum of the DyFeO\textsubscript{3} nanoparticles depicted in Fig.2(a) confirmed the presence of Dy, Fe, and O, consistent with the results from energy-dispersive X-ray (EDX) analysis. High-resolution XPS spectra of DyFeO\textsubscript{3} nanoparticles, presented in Fig.2(b-f), revealed the coexistence of Dy$^{4+}$, Dy$^{3+}$, Fe$^{2+}$, Fe$^{3+}$, and O$^{2-}$ ions \cite{liu2017synthesis,owusu2017low,yang2009characterization,liu2020oxygen}. These oxidation states might play the role of enhancing the charge storage capacity of the nanocomposite through a pseudocapacitive mechanism involving Faradaic redox reactions. The specific binding energies of all the XPS peaks for DyFeO\textsubscript{3} nanoparticles are documented in Table S2 in the supplementary information. Note that, in Fig.2(f), the high-resolution XPS spectra of O-1s were deconvoluted, yielding two distinct peaks at binding energies of 529.53 and 531 eV. These peaks correspond to metal-oxygen bonds and oxygen vacancies, respectively, with molar ratios of 61.33\% and 38.67\% \cite{liu2020oxygen}. These vacancies can arise from the charge imbalance resulting from the mixed oxidation states of rare earth ions (Dy\textsuperscript{3+} and Dy\textsuperscript{4+}) and transition metal ions (Fe\textsuperscript{2+} and Fe\textsuperscript{3+}) within the crystal lattice. The presence of oxygen vacancies in DyFeO\textsubscript{3} is likely to enhance charge carrier generation and improve charge transfer, potentially playing a promising role in enhancing the material's capacitance.


Notably, in the synthesis of DyFeO\textsubscript{3} perovskite nanoparticles using the sol-gel method, we introduced the chelating agent ethylene glycol to the solution to generate a polymeric metal cationic network that serves as the precursor for the gel. The thermal decomposition of ethylene glycol released gases including water vapor, carbon dioxide, and volatile organic compounds \cite{kistler1931coherent,al2003preparation}. These evolved gases generated internal pressure within the gel matrix, seeking paths of least resistance and ultimately leading to the formation of voids or pores. Concurrently, the decomposition induces thermal stress, initiating microcracks or voids in regions where internal pressure exceeds the material's structural strength \cite{kistler1931coherent}. Furthermore, the generated gas bubbles act as nucleation sites for pore formation, coalescing during the synthesis process to establish larger voids \cite{feinle2016sol}. It is noteworthy that the calcination process at 750 $^{\circ}$C also facilitated the removal of volatile components, such as organic residues and solvent remnants. This crucial process contributed significantly to the formation of the porous structure in DyFeO\textsubscript{3}. Moreover, during calcination, we exposed the material to high temperatures in a controlled atmosphere using N\textsubscript{2} gas. This process leads to the extraction of oxygen atoms from the lattice, resulting in the generation of oxygen vacancies. The porous morphology and the presence of these vacancies enhance the surface area of DyFeO\textsubscript{3}, thereby influencing its reactivity and functional properties. Collectively, these factors are essential to enhance the electrochemical performance of the  DyFeO\textsubscript{3} electrode material \cite{cao2021recent}. 


\begin{figure}[h!]%
\centering
\includegraphics[width=1\textwidth]{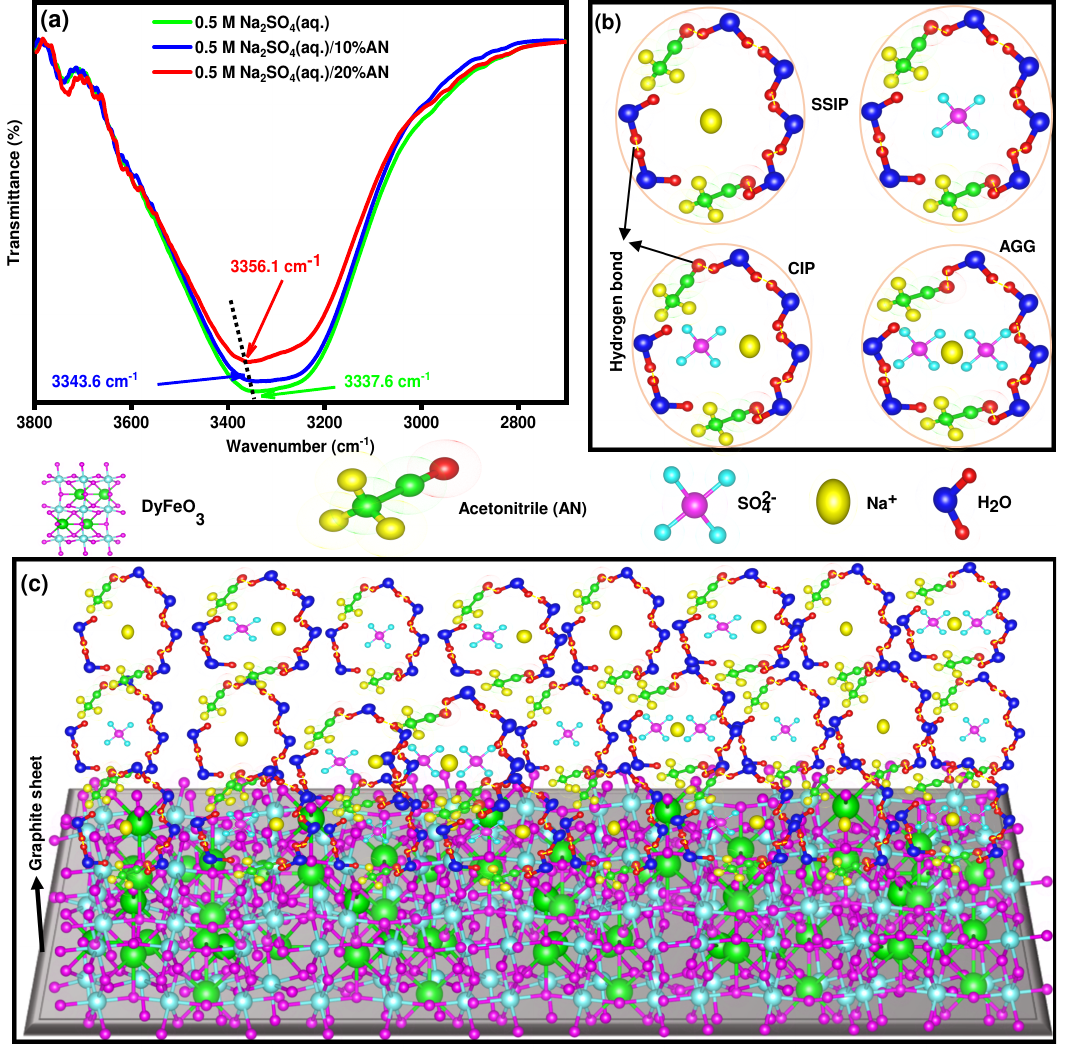}
\caption{\textbf{FTIR spectroscopy and solvation structure analysis were conducted on various electrolyte solutions to gain insights into their molecular interactions.} \textbf{a,} The FTIR spectra reveal a stretching mode in the range of 3337 to 3356 cm\textsuperscript{-1}, associated with hydrogen-bonded water molecules. The absence of a sharp decline in this stretching mode with changing acetonitrile (AN) concentration, coupled with a shift towards higher frequencies, suggests the dominance of hydrogen bonds in hybrid electrolyte solutions. This prevalent H-bonding diminishes the activity of free water molecules. \textbf{b,} SSIP, CIP and AGG solvation and coordination structure of ions (Na\textsuperscript{+} and $\text{SO}_4^{2-}$) in AN incorporated 0.5 M Na\textsubscript{2}SO\textsubscript{4}(aq.) electrolyte solution. \textbf{c,} A schematic illustration of the solvation structure and interfacial interaction between electrode and electrolyte ions in the AN incorporated 0.5 M Na\textsubscript{2}SO\textsubscript{4}(aq.) electrolyte solution. In this system, the synergistic effect of electrolyte components and electrode material significantly reduces the activity of free water molecules, which substantially enhances the ESW of the electrolyte solution.}\label{Fig_8}
\end{figure}

Before conducting electrochemical characterization, we examined the influence of various organic additives on the performance of water-in-salt (WIS) electrolytes. This was done through the incorporation of additives such as acetonitrile (AN) and ethylene glycol (EG) into a 0.5 M  Na\textsubscript{2}SO\textsubscript{4} aqueous electrolyte. The physical and chemical properties of the various electrolytes and solvent media were documented in Table S3 and Table S4. Comparative analysis revealed that, for enhancing the ESW of the WIS (0.5 M Na\textsubscript{2}SO\textsubscript{4}(aq.)) electrolyte, AN emerged as the most favorable organic additive. To elucidate the mechanism underlying the influence of AN additive on the ESW of the hybrid electrolytes, FT-IR was employed to verify the O–H bending and H-bending stretching vibration modes. In Fig. S14, the peak at 1641.64 cm\textsuperscript{-1} is attributed to the O–H bending vibration mode of water molecules \cite{huang2023ultrahigh}. An observable shift of the bending vibration peak from 1641.64 to 1645.64 cm\textsuperscript{-1} was noted with increasing AN ratio, indicative of enhanced O–H bonding, particularly in the 0.5 M Na\textsubscript{2}SO\textsubscript{4}(aq.)/20\%AN electrolyte solution \cite{huang2023ultrahigh}. In Fig. 3(a), the stretching mode at 3337.6 cm\textsuperscript{-1} corresponds to hydrogen-bonded water molecules. While a sharp decline in the stretching mode was not observed with varying AN concentrations but peak shifted from 3337.6 to 3356.1 cm\textsuperscript{-1}, indicating the dominance of hydrogen bonds in the electrolyte solutions \cite{huang2023ultrahigh}. These findings suggest that strengthened O–H bonding and hydrogen bonds contribute to the reduction in the activity of free water molecules and the inhibition of water decomposition. Moreover, AN decreases the viscosity of the electrolyte, with the bonding between AN and H\textsubscript{2}O identified as stronger than the bonds between H\textsubscript{2}O molecules. This phenomenon aids in the formation of a robust solid electrolyte interface (SEI) layer in 0.5 M Na\textsubscript{2}SO\textsubscript{4}(aq.)/20\%AN electrolyte solutions \cite{xu2022aqueous}.

The ion association interactions of AN incorporated 0.5 M Na\textsubscript{2}SO\textsubscript{4}(aq.) electrolyte solutions were investigated which are schematically shown in Fig. 3(b-c). According to the number of coordination bonds between anions and cations of electrolytes, the ion association interaction is generally divided into three categories \cite{huang2023ultrahigh,moon2022non}. First, solvent-separated ion pair (SSIP) involves solvation molecules coordinating with either Na\textsuperscript{+} or $\text{SO}_4^{2-}$ ions, keeping them separated within the solvent. This structure indicates a weaker ion association, as the ions are surrounded by solvent molecules, suggesting higher ion mobility. Second, contact ion pair (CIP) is characterized by a $\text{SO}_4^{2-}$ anion being coordinated directly to a Na\textsuperscript{+} cation, forming a stable structure with direct ion contact. This arrangement implies a moderate ion association strength, with the ions in closer proximity. The prevalence of CIP structures may indicate a more intimate interaction between Na\textsuperscript{+} and $\text{SO}_4^{2-}$, potentially influencing the kinetics of electrochemical reactions \cite{huang2023ultrahigh}. Third, anion–cation aggregate (AGG) refers to the coordination of $\text{SO}_4^{2-}$ anions with two or more Na\textsuperscript{+} ions or vice versa, forming larger aggregates or clusters. This structure suggests strong ion association and reduced solvation of ions within the aggregate, leading to decreased ion mobility. AGG structures may be more common in concentrated solutions or solvents with lower dielectric constants \cite{huang2023ultrahigh,moon2022non}. Conversely, in our electrolyte system, the SSIP and CIP structures were prevalent due to the dilute electrolyte and high dielectric constant of water, facilitating strong solvation capabilities through hydrogen bonding and dipole interactions. The addition of AN introduced hydrogen bonds with water molecules and reduced the activity of free water molecules. The coordination between solvent molecules and electrolyte ions weakened with increasing AN content, leading to a reduction in the decomposition of water molecules on the electrode surface. The low viscosity of AN facilitated ion mobility towards electrode surfaces, enhancing charge transport, while its relatively high dielectric constant contributed to the formation of a robust solvation layer \cite{xu2022aqueous}. These observed coordination structures were supported by FTIR spectra and confirmed the synergistic effect between solvent molecules and electrolyte ions which might be a key factor to improve ESW.

\begin{figure}[h]%
\centering
\includegraphics[width=1\textwidth]{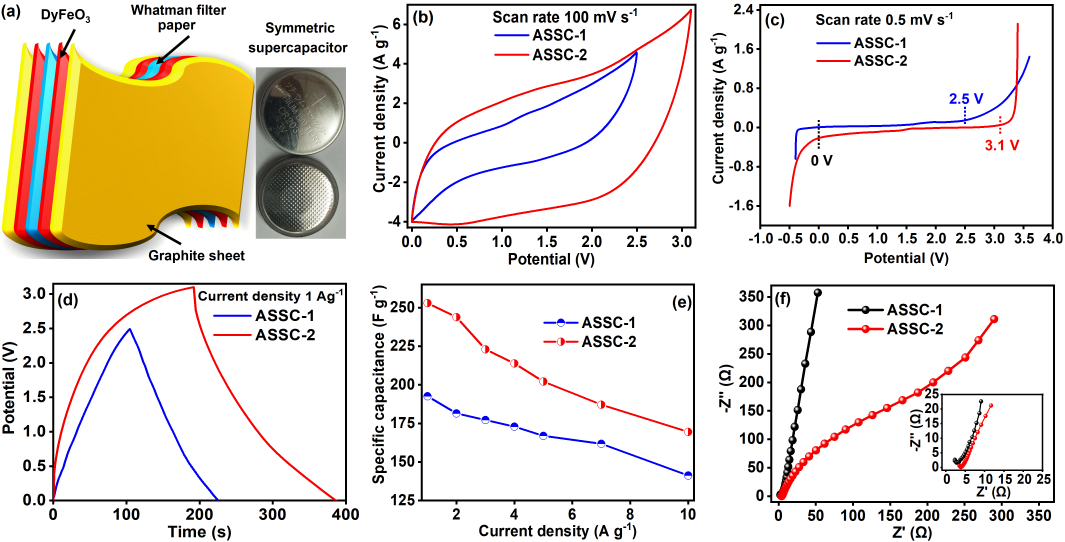}
\caption{\textbf{Electrochemical performances of the fabricated ASSCs (ASSC-1 and ASSC-2) using DyFeO\textsubscript{3} nanoparticles as an electrode material with 0.5 M Na\textsubscript{2}SO\textsubscript{4}(aq.) and 0.5 M Na\textsubscript{2}SO\textsubscript{4}(aq.)/20\%AN electrolyte solutions, graphite sheet as a current collector and Whatman filter paper as a separator}. \textbf{a,} Schematic and real illustration of the ASSCs. \textbf{b,} CV curves at a fixed scan rate of 100 mV s\textsuperscript{-1} show the large integrated CV area with ultra-high (3.1 V) potential window of ASSC-2 compared to ASSC-1. \textbf{c,} LSV curves demonstrate that the water oxidation potential of the ASSC-1 and ASSC-2 exceeds the theoretical water oxidation potential (1.23 V). \textbf{d,} GCD curves at a fixed current density of 1 A g\textsuperscript{-1}. \textbf{e,} At various current densities, the fabricated ASSCs exhibit high specific capacitance, \(C_{sp}\) upto 253.01 F g\textsuperscript{-1}. \textbf{f,} Nyquist plots show the very low impedance value i.e. low value of \(R_{s}\) and \(R_{ct}\), inset shows the enlarged view of the high-frequency region of Nyquist plots.}\label{Fig_9}
\end{figure}

\subsection{Electrochemical performance of as-fabricated aqueous symmetric supercapacitors (ASSCs)}
We aim to develop ASSCs using electrode materials with hybrid charge storage capability and electrolytes with a wide ESW, predominantly composed of water. In pursuit of our objectives, we initially conducted cyclic voltammetry (CV), linear sweep voltammetry (LSV), galvanostatic charge-discharge (GCD), and electrochemical impedance spectroscopy (EIS) analyses on the DyFeO\textsubscript{3} electrode material. These analyses were performed in various electrolytic solutions, including 0.5 M Na\textsubscript{2}SO\textsubscript{4}(aq.), 0.5 M Na\textsubscript{2}SO\textsubscript{4}(aq.)/10\%AN, 0.5 M Na\textsubscript{2}SO\textsubscript{4}(aq.)/10\%EG, 0.5 M Na\textsubscript{2}SO\textsubscript{4}(aq.)/20\%AN and 0.5 M Na\textsubscript{2}SO\textsubscript{4}(aq.)/20\%EG, using a three-electrode configuration. Detailed methodologies, along with corresponding outcomes and graphical representations, are extensively outlined in the ESI. We observed that even without additives, DyFeO\textsubscript{3} electrode material in a 0.5 M Na\textsubscript{2}SO\textsubscript{4}(aq.) electrolyte demonstrated exceptional electrochemical performance, featuring a notable ESW of 2.5 V, high specific capacitance (370.9 F g\textsuperscript{-1}), prolonged cycle life (94\%), and low impedance (\(R_{s}\) = 5.8 $\Omega$). Furthermore, the material exhibited elevated EDL capacitance, which is crucial for the development of high-power-density supercapacitors. Conversely, when subjecting the DyFeO\textsubscript{3} electrode material to electrochemical testing with an electrolyte solution comprising 0.5 M Na\textsubscript{2}SO\textsubscript{4}(aq.)/20\%AN, aimed at further increasing the ESW, the results revealed a widened ESW (3.1 V), increased capacitance (450 F g\textsuperscript{-1}), relatively prolonged cycle life (92\%), and a suitable equilibrium between the EDL and pseudocapacitance. 

In light of this favorable electrochemical performance, we fabricated the ASSCs using a coin cell arrangement to extend our investigation into the electrochemical properties of the DyFeO\textsubscript{3} electrode material in both 0.5 M Na\textsubscript{2}SO\textsubscript{4}(aq.) and 0.5 M Na\textsubscript{2}SO\textsubscript{4}(aq.)/20\%AN electrolyte solutions. To enhance clarity, for the remainder of this manuscript, the aqueous symmetric supercapacitor featuring DyFeO\textsubscript{3} electrode material and 0.5 M Na\textsubscript{2}SO\textsubscript{4}(aq.) electrolyte solution is denoted as ASSC-1, while the supercapacitor utilizing 0.5 M Na\textsubscript{2}SO\textsubscript{4}(aq.)/20\%AN electrolyte solution as ASSC-2. Initially, we conducted CV measurements of ASSC-1 at 2.5 V and ASSC-2 at 3.1 V across different scan rates to assess cyclic reversibility, as illustrated in supplementary Fig. S14(a-b). A substantial increase in the integrated CV area of ASSC-2, compared to ASSC-1, was observed, as shown in Fig. 4(b). This enhancement is attributed to the incorporation of 20\% AN in 0.5 M Na\textsubscript{2}SO\textsubscript{4}(aq.) electrolyte solution.

To assess the ESW of the ASSC-1 and ASSC-2, an LSV test was conducted at a constant scan rate of 0.5 mV s\textsuperscript{-1}, as illustrated in Fig. 4(c). The LSV curves reveal an impressive ultra-wide ESW of 2.5 V in additive-free ASSC-1 and 3.1 V in ASSC-2. Remarkably, minimal water splitting is observed within these potential windows. This broad electrochemical stability window can be ascribed to the unique characteristics of our synthesized electrode material, including diverse oxidation states, oxygen vacancies, and porous structure, along with the selected electrolyte composition featuring high conductivity, high dielectric constant, low viscosity, and effective solvation  \cite{cao2021recent}. The oxygen vacancies present in DyFeO\textsubscript{3} nanoparticles serve as adsorption sites for Na\textsuperscript{+} and $\text{SO}_4^{2-}$ ions within the electrolyte and influence the interaction between the electrode and electrolyte ion, which may provide alternative pathways for charge transfer that do not lead to water-splitting reactions. Moreover, the nano-size pores present in DyFeO\textsubscript{3} electrode facilitate efficient ion diffusion, as they are larger than the hydrated ion size of the electrolyte (as detailed in Table S3) \cite{vu2012porous}. This characteristic allows Na\textsuperscript{+} and $\text{SO}_4^{2-}$ ions to easily access the interior of the porous structure, enhancing the transport of charge carriers during the electrochemical process \cite{chen2022porous}. The synergistic effects of oxygen vacancies, the porous structure of DyFeO\textsubscript{3} nanoparticles, and the structure of the 0.5 M Na\textsubscript{2}SO\textsubscript{4}(aq.) electrolyte solution collectively contributed to achieving an ultra-wide ESW of ASSC-1. Conversely, the ESW of the ASSC-2 extends up to 3.1 V, attributed to the synergistic interplay of the DyFeO\textsubscript{3} electrode material and the solvation structure of 0.5 M Na\textsubscript{2}SO\textsubscript{4}(aq.)/20\%AN electrolyte solution, as discussed in the earlier section.

In Fig. 4(d), the comparative GCD curves of ASSC-1 and ASSC-2 exhibit that the charging-discharging duration of the ASSC-2 surpasses that of the ASSC-1. To a comprehensive understanding of the charging-discharging time under varied current densities (1, 2, 3, 4, 5, 7, and 10 A g\textsuperscript{-1}), the GCD curves of the ASSC-1 and ASSC-2 are presented in Fig. S15 (a,b). Notably, GCD curves of ASSC-1 exhibit a triangular shape, indicative of EDL capacitive-dominant behavior \cite{lei2013activated}. On the other hand, the GCD curves of ASSC-2 indicate pseudo-capacitive-dominant behavior. Moreover, the \(C_{sp}\) values were determined at various current densities using Eq. 1 based on the GCD curves, and the findings are depicted in Fig. 4(e) \cite{melchior2018high,tarek2023mos}. The ASSC-1 demonstrates a \(C_{sp}\) of 192.6 F g\textsuperscript{-1} at low current density (1 A g\textsuperscript{-1}) and 141.3 F g\textsuperscript{-1} at high current density (10 A g\textsuperscript{-1}), whereas for ASSC-2, the corresponding C\textsubscript{sp} values are 253 F g\textsuperscript{-1} and 169.5 F g\textsuperscript{-1}, respectively. In Fig. 4(f), the Nyquist plots reveal that the series resistance (\(R_{s}\)) of the ASSC-1 and ASSC-2 is 2.33 and 3.51 $\Omega$, respectively and the charge transfer resistance (\(R_{ct}\)) is 0.7 and 0.4 $\Omega$, respectively. The reduced impedance parameters observed can be ascribed to the presence of highly conductive small-sized ions (Na\textsuperscript{+} and $\text{SO}_4^{2-}$), along with the low viscosity and high dielectric constant of the solvent medium, which facilitates the easy penetration of electrolyte ions into the porous DyFeO\textsubscript{3} nanoparticles.

\begin{figure}[h]%
\centering
\includegraphics[width=1\textwidth]{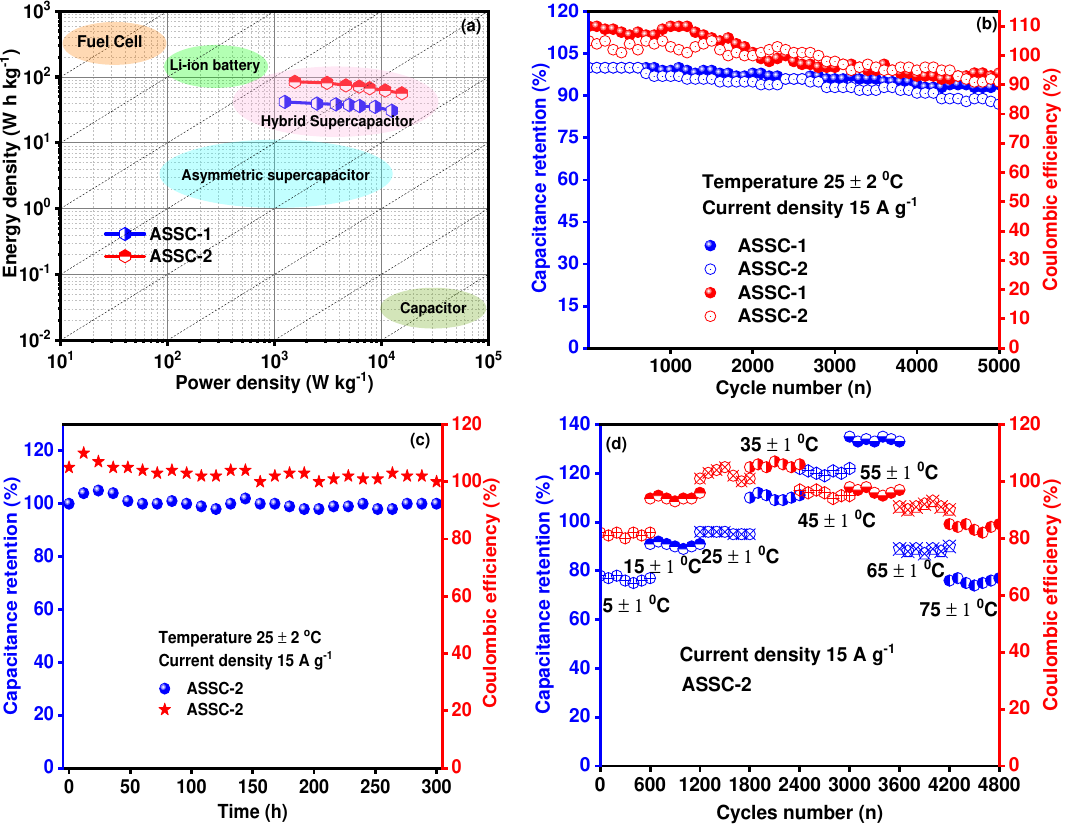}
\caption{\textbf{Ragone plot, specific capacitance retention and Coulombic efficiency of the ASSC-1 and ASSC-2}. \textbf{a,} Ragone plots exhibit high energy density comparable with Li-ion battery and high power density comparable with EDLC. \textbf{b,} At room temperature the fabricated aqueous symmetric supercapacitors exhibit impressive capacitance retention up to 81\% and Coulombic efficiency up to 94\% after 5000 GCD cycles, proving their durability over commercially available energy storage devices. \textbf{c,} The float voltage test also revealed superior capacitance retention and Coulombic efficiency of ASSC-2. \textbf{d,} At various temperatures (5 to 75 \textsuperscript{o}C) excellent capacitance retention and Coulombic efficiency are observed which is critically important for their practical application.}\label{Fig_10}
\end{figure}

The performance of energy storage devices is significantly influenced by their energy and power density, as these metrics play a crucial role in determining operational efficiency. To evaluate these key parameters, we calculated the energy and power density of the ASSCs using GCD data, employing Eq. 2 and Eq. 3 and the results are depicted in Fig. 5(a) through Ragone plot \cite{tarek2023mos}. Specifically, at a power density of 1250 W kg\textsuperscript{-1}, ASSC-1 shows an energy density of 41.81 W h kg\textsuperscript{-1}, and ASSC-2 exhibits an energy density of 84.43 W h kg\textsuperscript{-1} at a power density of 1550 W kg\textsuperscript{-1}. Remarkably, even at higher power densities, ASSC-1 and ASSC-2 maintain substantial energy density. For instance, at a power density of 12500 W kg\textsuperscript{-1}, ASSC-1 achieves an energy density of 30.65 W h kg\textsuperscript{-1}, and similarly, ASSC-2 exhibits an energy density of 56.57 W h kg\textsuperscript{-1} at 15500 W kg\textsuperscript{-1}. These findings highlight the excellent energy storage and delivery capabilities of our fabricated ASSCs.

Our fabricated ASSCs exhibit notable rate capability and durability at room temperature, as shown in Fig. 5(b), retaining 94\% and 87\% of the initial capacitance for ASSC-1 and ASSC-2, respectively. This performance is particularly significant for applications with frequent cycling, such as electric vehicles and renewable energy systems. Additionally, ASSC-1 and ASSC-2 exhibit Coulombic efficiencies of 94\% and 91\% respectively, indicating efficient charge-discharge processes and minimal energy losses. The observed high Coulombic efficiency is crucial for the overall performance of the energy storage devices, ensuring effective retrieval of a substantial proportion of stored energy during discharge.

The stability of the ASSC-2 was rigorously examined through the float voltage test, offering a more reliable assessment compared to repeated GCD cycling \cite{owusu2017low}. In this experiment, the assembled coin cell supercapacitor underwent a voltage application of 3.1 V, and a series of five charge and discharge cycles were carried out at a constant current density of 10 A g\textsuperscript{-1} with intervals of 12 hours, as illustrated in Fig. 5(c). The ASSC-2 exhibited remarkable stability over an extended testing period of 300 hours, maintaining approximately 100 $\pm$ 5 \% of its initial capacitance and Coulombic efficiency. These results demonstrate the robustness and efficiency of the ASSC-2 within a wide potential window of 3.1 V, which is essential for long-term stability under challenging operational conditions.

To investigate the practical applicability of our fabricated ASSC-2 across diverse temperature environments, the capacitance retention and Coulombic efficiency at high and low temperatures (5 to 75 $^{\circ}$C) were calculated as shown in Fig. 5(d). All calculations were conducted relative to results obtained at 25 $^{\circ}$C. The values of capacitance retention and Coulombic efficiency decreased at lower temperatures, with the lowest values observed at 5 $^{\circ}$C, being 75\% and 80\%, respectively, attributed to sluggish kinetics and diffusion processes \cite{masarapu2009effect,shafi2018enhanced}. Conversely, values increased up to 55 °C due to enhanced kinetics and efficient diffusion, reaching a maximum at this temperature of 135\% and 98\%, respectively. However, above 55 °C, both capacitance retention and Coulombic efficiency decreased, reaching their lowest points at 75 $^{\circ}$C —74\% and 82\%, respectively. This suggests that at higher temperatures, the advantages gained from improved kinetics are offset by heightened internal resistance. Despite these fluctuations, the ASSC-2 demonstrated superior durability and Coulombic efficiency across the wide temperature range, highlighting its promising practical applicability.

To assess the stability of DyFeO\textsubscript{3} electrode material following 5000 GCD cycles, once again we conducted XRD spectra, FESEM images, and electrochemical analyses (CV, GCD, and EIS) on the ASSC-2. The post-cycle characterizations provided in the Supplementary Information confirm the suitability of our synthesized DyFeO\textsubscript{3} nanoparticles as efficient electrode material for symmetric supercapacitors with a broad ESW.

The data in supplementary Table S7, including specific capacitance, ESW, capacitance retention, and energy density, demonstrate the superior performance of our fabricated ASSCs compared to other assembled supercapacitors. For instance, when tested in a costly 17 M NaClO\textsubscript{4}/H\textsubscript{2}O electrolyte solution, the AC//AC symmetric supercapacitor retains 66.1\% capacitance after 10,000 GCD cycles within a 2.3 V potential window \cite{bu2019low}. In contrast, ASSC-1, operated in a cost-effective and highly conductive 0.5 M Na\textsubscript{2}SO\textsubscript{4}(aq.) electrolyte solution with an ESW of 2.5 V, exhibits an impressive 94\% capacitance retention after 5000 GCD cycles.

Comparatively, a porous carbon symmetric supercapacitor, operating in a high-density 20 M LiTFSI WIS electrolyte solution with a wide 2.4 V ESW, exhibits a specific capacitance of 63 F g\textsuperscript{-1} at a current density of 0.5 A g\textsuperscript{-1}, accompanied by an energy density of 44 W h kg\textsuperscript{-1} at 564 W kg\textsuperscript{-1} \cite{xu2019temperature}. In contrast, our fabricated ASSC-2, featuring an extended ESW of 3.1 V, demonstrates a specific capacitance of 253.01 F g\textsuperscript{-1} and an energy density of 84.43 Wh kg\textsuperscript{-1} at a power density of 1550 W kg\textsuperscript{-1}, significantly surpassing the performance of numerous similarly assembled supercapacitors. In addition to its exceptional electrochemical properties, our fabricated electrolyte system (refer to Table S6) demonstrates a significantly lower production cost compared to many other electrolyte systems used previously. Although the cost of DyFeO\textsubscript{3} nanoparticles as an electrode material is slightly higher than that of commercially available carbon derivatives, the integrated production cost of our fabricated ASSCs might be more economical, considering their distinctive energy storage performances and the use of a low-cost, environmentally friendly electrolyte system.

\begin{figure}[h]%
\centering
\includegraphics[width=1\textwidth]{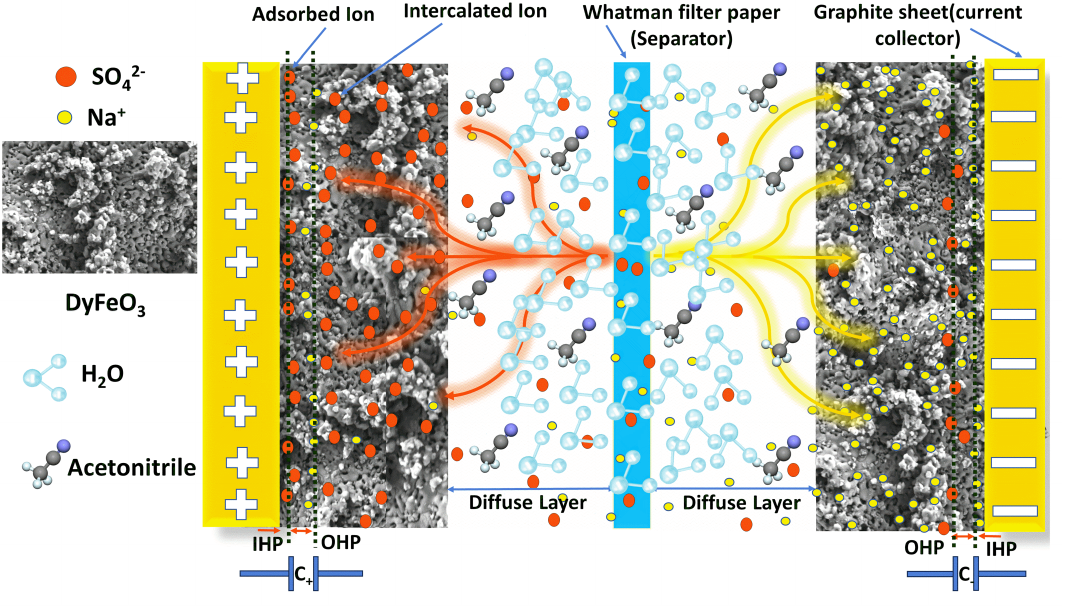}
\caption{\textbf{Schematic illustration of charge storage mechanism of the ASSC-2}. Upon the application of voltage, the electrolyte ionizes, producing Na\textsuperscript{+} and $\text{SO}_4^{2-}$ ions. These ions migrate toward the electrodes, forming an Inner Helmholtz Plane (IHP) adjacent to the electrode surface and an outer Helmholtz plane (OHP) extending into the bulk electrolyte. The interaction between the IHP and OHP defines the EDL architecture. This setup enables efficient charge separation, creating a high-capacitance region near the electrode surface. DyFeO\textsubscript{3} nanoparticles, known for their porous structures, play a crucial role by intercalating electrolyte ions. This intercalation enhances the system’s pseudocapacitive behavior and charge storage capacity. The electrolyte’s solvation structure, along with DyFeO\textsubscript{3}’s porous structure and oxygen vacancy, collectively contribute to increasing the electrochemical stability window to 3.1 V. Thus, the intricate charge storage mechanism and the energy storage potential of the system are demonstrated.}\label{Fig_12}
\end{figure}


The CV, GCD, and EIS analyses of the ASSCs illustrate the combined influence of EDL and pseudo-capacitive mechanisms on overall charge storage, as schematically depicted in Fig. 6 and further explained in the supplementary information. In this context, the EDL, composed of IHP and OHP, forms through electrostatic charge separation at the electrode-electrolyte interfaces. Simultaneously, the pseudo-capacitance results from either electrolyte ion intercalation into the DyFeO\textsubscript{3} electrode surface or redox interactions between DyFeO\textsubscript{3} nanoparticles and electrolyte ions. In Fig. S18, our fabricated ASSC-2 effectively illuminated an LED,  showcasing the real-world utility of DyFeO\textsubscript{3} as a hybrid electrode material for high-voltage ASSCs.

In conclusion, we present the first demonstration of a 2.5 V aqueous symmetric supercapacitor utilizing single-structured DyFeO\textsubscript{3} both as cathode and anode. The porous DyFeO\textsubscript{3} nanostructure emerges as a promising supercapacitor electrode material for practical applications, offering simplicity in preparation, environmental friendliness, and favorable electrochemical performance across a wide potential window in pure aqueous and aqueous dominating electrolytes. Incorporating 20\% acetonitrile in a 0.5 M Na\textsubscript{2}SO\textsubscript{4} aqueous electrolyte expanded the electrochemical stability window of the fabricated symmetric supercapacitor to 3.1 V, surpassing that of common aqueous-dominating ones. The superior performance of this hybrid supercapacitor can be attributed to the synergistic interplay of oxygen vacancies, the porous nanostructure, varied oxidation states of the constituent elements of DyFeO\textsubscript{3}, and the solvation structure of the electrolytes. The as-fabricated hybrid supercapacitor displays a high capacitance of 253 F g\textsuperscript{-1} at a current density of 1 A g\textsuperscript{-1}, an impressive energy density of 84.43 Wh kg\textsuperscript{-1} at a power density of 1550 W kg\textsuperscript{-1} in 0.5 M Na\textsubscript{2}SO\textsubscript{4}(aq.)/20\%AN electrolyte, maintaining 87\% capacitance retention, and 92\% Coulombic efficiency after 5000 cycles. This significant advancement paves the way for the development of the next generation of high-performance energy storage devices based on a porous rare-earth orthoferrite electrode material in aqueous or aqueous-dominating electrolyte solutions.

\bibliography{sn-bibliography}

\section{Acknowledgements} 
We sincerely acknowledge the University Grant Commission of Bangladesh for financial support, which has been instrumental in the successful execution of this research. We also acknowledge Bangladesh University of Engineering and Technology for providing essential resources crucial for conducting our experiments.

\section{Competing financial interests}
The authors declare no competing financial interests.

\section{Methods}
\subsection{Synthesis of DyFeO\textsubscript{3} nanoparticles}
Nanoparticles of the DyFeO\textsubscript{3} perovskite were synthesized using a sol-gel technique, as illustrated in Fig. S1 \cite{ansari2019physico}. Initially, stoichiometric amounts of Dy(NO\textsubscript{3})\textsubscript{3}·5H\textsubscript{2}O (Sigma-Aldrich, 99.80\%) and Fe(NO\textsubscript{3})\textsubscript{3}·9H\textsubscript{2}O (Sigma-Aldrich, 98.00\%) were individually dissolved in 100 ml of deionized water and stirred for 15-20 minutes using a magnetic stirrer. Subsequently, the solutions were combined, and citric acid (C\textsubscript{6}H\textsubscript{8}O\textsubscript{7}) was introduced into the mixture. citric acid served as a chelating agent, forming stable complexes with the metal ions. The solution's pH was adjusted to 7 by incorporating NH\textsubscript{4}OH. Subsequently, the chelating agent, ethylene glycol, was introduced to generate a polymeric metal cationic network, serving as the precursor for the gel. After four hours, the temperature was raised to 200 $^{\circ}$C, leading to the complete combustion of the gel and the formation of the desired powder. The combustion process entails the decomposition of the organic components within the gel, liberating gases and leaving behind metal oxides. The resulting material was finely grounded using an agate mortar. To achieve the desired crystallization and porous structure, the material underwent calcination at 750 $^{\circ}$C for 6 hours, with a heating rate of 5 $^{\circ}$C per minute in a nitrogen environment. The controlled heating rate of 5 $^{\circ}$C per minute minimizes thermal stress, facilitating controlled decomposition. Nitrogen gas flow, particularly during the initial stages of the calcination process, ensures an oxygen-free environment. This controlled atmosphere is crucial for preventing unwanted oxidation reactions and influencing the evolution of gases during decomposition. The reproducible formation of porosity in DyFeO\textsubscript{3} was confirmed through the judicious selection of solvent (here water), systematic adjustment of gel precursor concentration (here ethylene glycol), careful control of temperature and optimization of the calcination temperature in a N\textsubscript{2} environment.

\subsection{Characterization methods and instrumentation}
\subsubsection{Structural and morphological characterization}
The crystal structure of the synthesized materials was determined using a Rigaku SmartLab X-ray diffractometer with copper K-Alpha radiation. Further insights were gained through a thorough analysis of the powder X-ray diffraction (XRD) patterns, employing the Rietveld refinement method with the Fullprof software program. The surface morphology was analyzed using both Field Emission Scanning Electron Microscopy (FESEM) and Transmission Electron Microscopy (TEM). Elemental insights were obtained through energy-dispersive X-ray (EDX) spectra, meticulously revealing the elemental composition of the synthesized samples. Fourier Transform Infrared Spectroscopy (FTIR) was employed for a comprehensive examination of chemical bonds and functional groups. X-ray Photoelectron Spectroscopy technique revealed the chemical states and binding energies inherent in the DyFeO\textsubscript{3}. 

\subsubsection{Electrochemical Characterizations}
Cyclic voltammetry (CV), linear sweep voltammetry (LSV), galvanostatic charge-discharge (GCD), and electrochemical impedance spectroscopy (EIS) analyses were systematically performed on the DyFeO\textsubscript{3} electrode material across diverse electrolyte solutions. These assessments were conducted using both three-electrode and symmetric two-electrode configurations, including our fabricated coin cell supercapacitor. The experiments were carried out utilizing an electrochemical workstation (Metrohm Autolab PGSTAT302N). The detailed procedure for preparing the working electrode slurry is outlined in the Supplementary Information, and Fig. S2 schematically illustrates both the preparation process of the working electrode slurry and the electrochemical setup of the three-electrode system.

\textbf{Assembly of coin cell supercapacitors in symmetric configuration:}
Supplementary Fig. S3 depicts the configuration of the CR2032 coin cell and the corresponding fabrication steps. In the context of the coin cell supercapacitor, DyFeO\textsubscript{3} nanoparticles served as the electrode material for both the anode and cathode. The electrolyte comprised of 0.5 M Na\textsubscript{2}SO\textsubscript{4}(aq.) and 0.5 M Na\textsubscript{2}SO\textsubscript{4}(aq.)/20\%AN solutions, Whatman filter paper functioned as the separator, and disc-shaped graphite sheets served as current collectors for both the anode and cathode. In the coin cell assembly, a slurry of the active material was uniformly applied to two individual graphite sheets (diameter 1.65 cm) at a consistent loading of 10 mg cm\textsuperscript{-2}. The treated graphite sheets, designated as the anode and cathode, were then dried in an oven at 80 \textsuperscript{o}C for 12 hours. A circular Whatman filter paper, saturated for 24 hours with 0.5 M Na\textsubscript{2}SO\textsubscript{4}(aq.) and 0.5 M Na\textsubscript{2}SO\textsubscript{4}(aq.)/20\%AN electrolyte solutions, was positioned between the modified graphite sheets to finalize the coin cell arrangement. Subsequently, the coin cell supercapacitor was sealed under a pressure of 800 kg using an automatic coin cell assembling machine (TMAX-160S).

\subsection{Mathematical analysis}
\subsubsection{Electrochemical Measurement}
\textbf{Specific capacitance (\(C_{sp}\))}

The specific capacitance is computed using the subsequent equation tailored for the symmetric two-electrode system (detailed derivation is provided in the supplementary information).
 \cite{melchior2018high,tarek2023mos}.
\begin{equation}
    C_{sp}=\frac{4J\Delta{t}}{\Delta{V}}
    \label{eq6}
\end{equation}
Where, J, $\Delta{t}$ and $\Delta{V}$ stand for current density, time to discharge and potential window, respectively.

\textbf{Energy density (\(E_{sp}\)) and Power density (\(P_{sp}\))}

The specific energy density (\(E_{sp}\), measured in W h kg\(^{-1}\)), and specific power density (\(P_{sp}\), measured in W kg\(^{-1}\)) for the symmetric two-electrode system (pertaining to a single electrode) were computed based on the GCD using Equations 4 and 5, respectively \cite{melchior2018high,tarek2023mos}.
\begin{equation}
    E_{sp}=\frac{C_{sp}(\Delta{V})^2\times1000}{8\times3600}
    \label{eq7}
\end{equation}
\begin{equation}
    P_{sp}=\frac{E_{sp}\times3600}{\Delta{t}}
    \label{eq8}
\end{equation}

\textbf{Capacitance retention (\%)}
\begin{equation}
\text{Capacitance retention (\%)} = \left( \frac{\text{Capacitance at } n^\text{th} \text{ cycles}}{\text{Initial capacitance}} \right) \times 100\%
\end{equation}
\textbf{Coulombic efficiency (\%)}
\begin{equation}
\text{Coulombic efficiency (\%)} = \left( \frac{\text{Discharging time}}{\text{Charging time}} \right) \times 100\%
\end{equation}

\end{document}